\documentclass[showkeys,onecolumn]{revtex4}
\usepackage{amsmath}
\usepackage{amscd}
\usepackage{graphicx}
\usepackage{subfigure}
\usepackage{appendix}
\usepackage{amsfonts}
\usepackage{multirow}
\usepackage{bm}
\usepackage{bbm}

\begin{document}

\title[Short Title]{Accelerating population transfer in a transmon qutrit via Shortcuts to adiabaticity}

\author{Ye-Hong Chen$^{1,2}$}
\author{Zhi-Cheng Shi$^{1,2,}$\footnote{E-mail: szc2014@yeah.net}}
\author{Jie Song$^{3}$}
\author{Yan Xia$^{1,2,}$\footnote{E-mail: xia-208@163.com}}
\author{Shi-Biao Zheng$^{1,2}$}

\affiliation{$^{1}$Department of Physics, Fuzhou University, Fuzhou 350002, China\\
             $^{2}$Fujian Key Laboratory of Quantum Information and Quantum Optics (Fuzhou University), Fuzhou 350116, China\\
             $^{3}$Department of Physics, Harbin Institute of Technology, Harbin 150001, China}


\begin{abstract}
  In this paper, a method to accelerate population transfer by designing nonadiabatic evolution paths is proposed.
  We apply the method to realize robust and accelerated population transfer with a transmon qutrit.
  By numerical simulation, we show that this method allows a robust population transfer between the ground states in a $\Lambda$ system.
  Moreover, the total pulse area for the population transfer is low as $1.9\pi$ that verifies the evolution is accelerated without increasing the pulse intensity.
  Therefore, the method is easily implementable based on the modern pulse shaper technology and it provides selectable schemes
  with interesting applications in quantum information processing.
\end{abstract}


\maketitle
\section{Introduction}
Coherent control of the quantum state is a critical element for various quantum technologies such as high-precision
measurement \cite{Rmp781297}, coherent manipulation of atom and
molecular systems \cite{Rmp7953}, and quantum information processing \cite{WVB08,Prl95080502}.
Recently, an increasing interest has been devoted to study an approach named ``Shortcuts to adiabaticity'' (STA) which aims at designing nonadiabatic
methods to accelerate the adiabatic process \cite{Prl105123003,ETSISMGMMACDGOARXCJGMAmop13,Prl111100502,Prl109100403,Jpca1079937,Jpa42365303,Prsa4661135,Oc13948,Epjst224189,Pra86033405,Pra89053408,Pra89043408,Nc712999,
Pra82033430,Epl9323001,Pra84031606Epl9660005,Jpb43085509,Pra95062319,Pra93052324,Prl116230503,Pra93052109,Natc712479,Pra84013428}.
By applying STA, one can drive a quantum system from a given initial state to a prescribed final state in a shorter time than adiabatic process
without losing its robustness property \cite{ETSISMGMMACDGOARXCJGMAmop13}.
There are now a rich set of STA techniques devoted to speed up slow adiabatic processes,
such as, counterdiabatic driving \cite{Jpca1079937,Jpa42365303}, invariant-based inverse engineering \cite{Pra86033405,Pra89053408,Pra89043408},
fast-forward scaling \cite{Prsa4661135}, multiple Schr\"{o}dinger dynamics \cite{Prl109100403,Pra93052324}, dressed-state-based shortcuts \cite{Prl116230503}.
Generally speaking, according to the differences of evolution paths, the STA techniques fall into two major categories: 
(i) The Hamiltonian $H(t)$ is constructed to make the dynamics adiabatic with respect to a reference Hamiltonian $H_{0}(t)$; (ii) $H(t)$ is constructed without making explicit use of a reference Hamiltonian  $H_{0}(t)$.
Counterdiabatic driving proposed by Rice and Demirplak \cite{Jpca1079937} or Berry \cite{Jpa42365303} is a typical example
of the (i)-type STA technique. The principle is by using a supplementary Hamiltonian to suppress
transitions between different time-dependent instantaneous eigenstates (adiabatic basis) of a reference time-dependent Hamiltonian.
In this way, each of the instantaneous eigenstates of $H_{0}(t)$ can evolve along itself all the time without the requirement of adiabatic condition so that the evolution speed is improved.
However, the designed supplementary Hamiltonian is usually hard to realize in practice.
Invariant-based inverse engineering is a typical example
of the (ii)-type STA technique. The evolution path is given based on the eigenstates of the system's invariant rather than the reference Hamiltonian $H_{0}(t)$.
The possible difficulty in applying this kind of STA technique is that finding invariants for an arbitrary Hamiltonian is still a challenge.
Similar difficulties also exist in other (ii)-type STA techniques \cite{Pra86033405,Pra89053408,Pra89043408,Prsa4661135,Prl109100403,Pra93052324,Prl116230503,Jpb43085509,
Pra93052109,Pra95062319} that the nonadiabatic evolution paths are hard to be found.

In this paper, we focus on improving the (ii)-type STA techniques.
By reverse engineering, we come up with an idea to search for the desired nonadiabatic evolution paths.
The strategy is to design a time-dependent
vector $|\phi_{0}(t)\rangle=\sum_{n}A_{n}|n\rangle$ as the nonadiabatic evolution path,
where $|n\rangle$ are the eigenstates of the identity matrix $\bm{1}$ (totally time-independent)
and $A_{n}$ are the probability
amplitudes of $|n\rangle$ satisfying $\sum_{n}|A_{n}|^{2}=1$.
Then, we accordingly write down the orthogonal partners of $|\phi_{0}(t)\rangle$ to form a complete Hilbert space.
If the vector $|\phi_{0}(t)\rangle$ is designed to be decoupled with each of its
orthogonal partners from beginning to end, 
the evolution of the system will exactly follow the time-dependent
vector $|\phi_{0}(t)\rangle$ when the system is initially in $|\phi_{0}(t)\rangle$ \cite{Prl581593,Prl602339}.
To realize this idea, the key point is to find the analytical orthogonal partners for the path $|\phi_{0}(t)\rangle$.
We present an accepted way to analytically construct orthogonal complete vectors in arbitrary-dimension
space in Sec. II. The starting point is a two-dimension orthogonal complete basis formed by
trigonometric functions with angle $\theta_{1}$.
Then, with a series of simple unitary matrixes $\bm{A}^{(k)}$ which are formed by trigonometric functions with angle $\theta_{k}$,
a set of orthogonal complete vectors $|\zeta_{n}\rangle$ will be constructed by relationship $|\zeta_{n}\rangle=\sum_{m}A_{n,m}|m\rangle$,
where $\bm{A}=\prod_{k}\bm{A}^{(k)}$ and $A_{n,m}$ are the matrix elements of unitary matrix $\bm{A}$.
Hence, one of the vectors $|\zeta_{n}\rangle$ can be chosen as the evolution path
and the others can be accordingly chosen as its orthogonal partners.
In the rotating frame, the condition to decouple the evolution path $|\phi_{0}(t)\rangle$ from its orthogonal partners can be analytically solved: its is a set of linear equations as given in Eq. (\ref{eq1-6}).
Beware that, in order to ensure the Eq. (\ref{eq1-6}) is analytically solvable, the designed evolution path is better to satisfy the following two points:
(i) the functional form of the evolution path $|\phi_{0}(t)\rangle$
should be more simple than that of its orthogonal partners;
(ii) the elements in the evolution path $|\phi_{0}(t)\rangle$ are better to be nonzero.

As an application example, we apply the present method to perform a robust and accelerated population transfer within
a transmon-type qutrit. A qutrit, constituted by
the lowest three levels of the system, can be coupled to the
microwave drivings, consisting of ac gate voltage and timedependent
bias flux. Allowed by the level-transition rule,
we address a $\Lambda$-configuration interaction. By applying the present method, the population transfer
can be accelerated remarkably in contrast with the
adiabatic operation as demonstrated by numerical simulation. We also analyze
the total pulse area which is used to measure the total energy cost for the accelerated process.
The result shows the present method allows the robust
population transfer between the ground states in a $\Lambda$ system with the total pulse area as low as $1.9\pi$. Such a pulse
area is small enough to verify that the population transfer is accelerated.

The rest of the paper is structured as follows.
The general method to construct orthogonal complete basis is given
in Sec. II. The general condition to decouple a vector from its orthogonal
partners is given in Sec. III. In Sec. IV, transferring
population with negligible leakages can be drastically sped up within a qutrit.
In Sec. V, we check the system's robustness against systematic errors and amplitude-noise errors.
In Sec. VI, we give the conclusion.

\section{Constructing orthogonal complete time-dependent vectors in high-dimension space}
Finding analytical eigenstates for a general
Hamiltonian does not have a tractable algorithm,
but finding an arbitrary set of orthogonal complete basis for a given dimension
Hilbert space is a much easier work.
We know, for an $N$-dimension Hilbert space, $|n\rangle$ ($n=1,2,3,\cdots,N$)
is a natural set of orthogonal complete basis.
A set of orthogonal complete basis $|\zeta_{n}\rangle$ can be obtained by
performing an orthogonal transformation on $|m\rangle$ ($m=1,2,3,\cdots,N$) as $|\zeta_{n}\rangle=\sum_{m}A_{n,m}|m\rangle$,
where $A_{n,m}$ are the matrix elements of unitary matrix $\bm{A}$.
The $N$-dimension unitary matrix $A$ can be in fact obtained
by $\bm{A}=\prod_{k}\bm{A}^{(k)}$ with $\bm{A}^{(k)}$ being a series of 
unitary matrixes. 

For example, $\bm{A}^{(1)}=\left(
                        \begin{array}{cc}
                          \cos\theta_{1} & e^{i\chi_{1}}\sin{\theta_{1}} \\
                          \sin\theta_{1} & -e^{i\chi_{1}}\cos{\theta_{1}} \\
                        \end{array}
                      \right)
$
is one of orthogonal matrixes in two-dimension Hilbert space. Then the orthogonal complete basis $|\zeta^{(1)}_{n}\rangle$ can be constructed as
\begin{align}\label{eqa-1}
  |\zeta_{1}^{(1)}\rangle=&\cos{\theta_{1}}|1\rangle+e^{i\chi_{1}}\sin{\theta_{1}}|2\rangle,\cr
  |\zeta_{2}^{(1)}\rangle=&\sin{\theta_{1}}|1\rangle-e^{i\chi_{1}}\cos{\theta_{1}}|2\rangle.
\end{align}
If we choose $\bm{A}^{(2)}$ with a similar form as $\bm{A}^{(1)}$ but different parameters, say,
$\bm{A}^{(2)}=\left(
                        \begin{array}{cc}
                          \cos\theta_{2} & e^{i\chi_{2}}\sin{\theta_{2}} \\
                          \sin\theta_{2} & -e^{i\chi_{2}}\cos{\theta_{2}} \\
                        \end{array}
                      \right)
$,
another set of orthogonal vectors $|\zeta^{(2)}_{n}\rangle=\sum_{m}(\bm{A}^{(2)}\bm{A}^{(1)})_{n,m}|m\rangle$ are obtained as
\begin{align}\label{eqa-1b}
  |\zeta_{1}^{(2)}\rangle=&\left(
                               \begin{array}{c}
                                 \cos{\theta_{1}}\cos{\theta_{2}}+e^{i\chi_{2}}\sin{\theta_{1}}\sin{\theta_{2}} \\
                                 e^{i\chi_{1}}(\sin{\theta_{1}}\cos{\theta_{2}}-e^{i\chi_{2}}\cos{\theta_{1}}\sin{\theta_{2}})
                               \end{array}
                          \right), \cr
  |\zeta_{2}^{(2)}\rangle=&\left(
                               \begin{array}{c}
                                 \cos{\theta_{1}}\sin{\theta_{2}}-e^{i\chi_{2}}\sin{\theta_{1}}\cos{\theta_{2}} \\
                                 e^{i\chi_{1}}(\sin{\theta_{1}}\sin{\theta_{2}}+e^{i\chi_{2}}\cos{\theta_{1}}\cos{\theta_{2}})
                               \end{array}
                          \right).
\end{align}
It is still a set of two-dimension vectors.

In higher-dimension space, the expression for the unitary matrix $\bm{A}^{(1)}$ is assumed to be a common form. For example,
in the three-dimension space, $\bm{A}^{(1)}$ is expressed as
$\bm{A}^{(1)}=\left(
  \begin{array}{ccc}
    \cos\theta_{1} & e^{i\chi_{1}}\sin{\theta_{1}} & 0 \\
    \sin\theta_{1} & -e^{i\chi_{1}}\cos{\theta_{1}} & 0 \\
    0 & 0 & 1 \\
  \end{array}
\right)$,
in the four-dimension space, $\bm{A}^{(1)}$ is expressed as
$\bm{A}^{(1)}=\left(
  \begin{array}{cccc}
    \cos\theta_{1} & e^{i\chi_{1}}\sin{\theta_{1}} & 0 & 0\\
    \sin\theta_{1} & -e^{i\chi_{1}}\cos{\theta_{1}} & 0 & 0\\
    0 & 0 & 1 & 0  \\
    0 & 0 & 0 & 1 \\
  \end{array}
\right)$,
and so on.
The $\bm{A}^{(k)}$ ($k>1$) are given by exchanging the rows or columns of $\bm{A}^{(1)}$. For example,
$\bm{A}^{(2)}$ and $\bm{A}^{(3)}$ in a three-dimension space can be given as
$\bm{A}^{(2)}=\left(
  \begin{array}{ccc}
    0 & e^{i\chi_{2}}\sin{\theta_{2}} & \cos\theta_{2} \\
    0 & -e^{i\chi_{2}}\cos{\theta_{2}} & \sin\theta_{2}  \\
    1 & 0 & 0 \\
  \end{array}
\right)$
and
$\bm{A}^{(3)}=\left(
  \begin{array}{ccc}
    0 & e^{i\chi_{3}}\sin{\theta_{3}} & \cos\theta_{3} \\
    0 & -e^{i\chi_{3}}\cos{\theta_{3}} & \sin\theta_{3}  \\
    1 & 0 & 0 \\
  \end{array}
\right)$,
respectively.
A common set of orthogonal vectors $|\zeta^{(3)}_{n}\rangle=\sum_m(\bm{A}^{(3)}\bm{A}^{(2)}\bm{A}^{(1)})_{n,m}|m\rangle$ in a three-dimension space are thus constructed as
\begin{align}\label{eqa-5}
  |\zeta_{1}^{(3)}\rangle=&\left(
                                \begin{array}{c}
                                  \cos{\theta_{1}}\cos{\theta_{3}}-e^{i(\chi_{2}+\chi_{3})}\sin{\theta_{1}}\cos{\theta_{2}}\sin{\theta_{3}} \\
                                  e^{i\chi_{1}}[\sin{\theta_{1}}\cos{\theta_{3}}+e^{i(\chi_{2}+\chi_{3})}\cos{\theta_{1}}\cos{\theta_{2}}\sin{\theta_{3}}] \\
                                  e^{i\chi_{3}}\sin{\theta_{2}}\sin{\theta_{3}}
                                \end{array}
                          \right),\cr
  |\zeta_{2}^{(3)}\rangle=&\left(
                                \begin{array}{c}
                                  \cos{\theta_{1}}\sin{\theta_{3}}+e^{i(\chi_{2}+\chi_{3})}\sin{\theta_{1}}\cos{\theta_{2}}\cos{\theta_{3}} \\
                                  e^{i\chi_{1}}[\sin{\theta_{1}}\sin{\theta_{3}}-e^{i(\chi_{2}+\chi_{3})}\cos{\theta_{1}}\cos{\theta_{2}}\cos{\theta_{3}}] \\
                                  -e^{i\chi_{3}}\sin{\theta_{2}}\cos{\theta_{3}}
                                \end{array}
                          \right),\cr
  |\zeta_{3}^{(3)}\rangle=&\left(
                                \begin{array}{c}
                                  e^{i\chi_{2}}\sin{\theta_{1}}\sin{\theta_{2}} \\
                                  -e^{i(\chi_{1}+\chi_{2})}\cos{\theta_{1}}\sin{\theta_{2}} \\
                                  \cos{\theta_{2}}
                                \end{array}
                          \right),
\end{align}
which satisfy the condition (ii) the elements in the evolution path $|\phi_{0}(t)\rangle$ are nonzero.

\section{The general method to decouple a vector from its orthogonal partners with a given Hamiltonian}
We start from the Schr\"{o}dinger equation for a quantum system
\begin{align}\label{eq1-1}
  i\hbar\partial_{t}|\psi_{0}(t)\rangle=H_{0}(t)|\psi_{0}(t)\rangle,
\end{align}
where $\partial_{t}\equiv\frac{\partial}{\partial t}$.
As we know, the general solution for the non-linear equation in Eq. (\ref{eq1-1}) can be expressed as
\begin{align}\label{eq1-2}
  |\psi_{0}(t)\rangle=\sum_{n}C_{n}(t)|\phi_{n}(t)\rangle,
\end{align}
where $C_{n}(t)$ are time-dependent coefficients and $|\phi_{n}(t)\rangle$ are a set of orthogonal time-dependent vectors satisfying
\begin{align}\label{eq1-3}
  \langle\phi_{n}(t)|\phi_{m}(t)\rangle=\delta_{n,m},\
  \sum_{n}|\phi_{n}(t)\rangle\langle\phi_{n}(t)|=\bm{1}.
\end{align}
In order to study the dynamical evolution of the system, we accordingly define picture rotation matrixes
\begin{align}\label{eq1-4}
  R^{\dag}=\sum_{n}|\phi_{n}(t)\rangle\langle n|,\  \text{and}\ R=\sum_{n}|n\rangle\langle\phi_{n}(t)|,
\end{align}
where $|n\rangle$ are the eigenstates of the identity matrix $\bm{1}$.
Then, the state in the rotating frame becomes $|\psi_{1}(t)\rangle=R|\psi_{0}(t)\rangle$.
In this case, the dynamical evolution after picture transformation is described as
$i\hbar\partial_{t}|\psi_{1}(t)\rangle=H_{1}(t)|\psi_{1}(t)\rangle$, where
\begin{align}\label{eq1-5}
  H_{1}(t)=RH_{0}(t)R^{\dag}-i\hbar R(\partial_{t}{R^{\dag}}).
\end{align}

Here, we would like to emphasize that the off-diagonal terms
represent the couplings between vectors $|\phi_{n}(t)\rangle$.
If we choose $\langle m|H_{1}(t)|0\rangle=0$ ($m\neq 0$), the vector $|\phi_{0}(t)\rangle$
will be decoupled to $|\phi_{m}(t)\rangle$ \cite{Prsla42961}. Which means, the time-dependent vector $|\phi_{0}(t)\rangle$
will evolve along itself all the time without transition to $|\phi_{m}(t)\rangle$.
In this case, we have $|C_{0}(t)|=|C_{0}(t_{i})|$.
Then, if we choose $|C_{0}(t_{i})|=1$ and $|C_{m\neq 0}(t_{i})|=0$ (the system is initially in $|\phi_{0}(t)\rangle$),
the system will be ensured in $|\phi_{0}(t)\rangle$ all the time without transition to others.
As long as the condition
\begin{align}\label{eq1-6}
  \langle 0|H_{1}(t)|m\rangle=\langle m|H_{1}(t)|0\rangle=0,\ \ \ (m\neq 0)
\end{align}
is satisfied, one can drive the system from a given initial state to a prescribed final state in a shorter time through a nonadiabatic path $|\phi_{0}(t)\rangle$.

\begin{figure}
 \scalebox{0.3}{\includegraphics {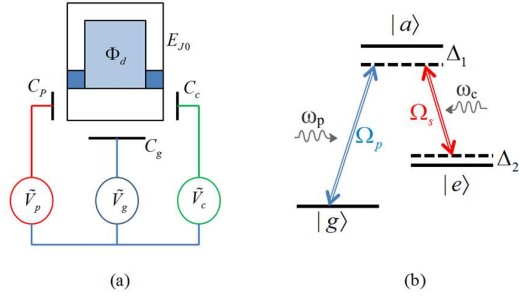}}
 \caption{
         (a) Schematic diagram of the considered artificial atom driven by a control microwave field $\tilde{V}_{c}$ and a probe field $\tilde{V}_{p}$.
         (b) For a $\Lambda$-type interaction, the control (probe) field with frequency $\omega_{c}$ ($\omega_{p}$) aims at causing the transition between $|a\rangle$ and $|e\rangle$ ($|g\rangle$).
         }
 \label{fig0}
\end{figure}

\section{Accelerated population transfer in A TRANSMON-TYPE ARTIFICIAL ATOM
WITH WEAK LEVEL ANHARMONICITY}
We consider a transmon-type Cooper-pair box (CPB) circuit which contains a
superconducting box with $n$ extra Cooper pairs.
The CPB is connected to a segment of a superconducting loop through two symmetric Josephson junctions with
the identical coupling energies $E_{J0}$.
A static gate voltage $V_{g}$ applied to the gate capacitance $C_{g}$ induces offset charges. A magnetic
flux $\Phi_{d}$ threading the loop is used to modulate the effective
Josephson coupling, $E_{J}=2E_{J0}\cos{(\pi\Phi_{d}/\Phi_{0})}$, where $\Phi_{0}=\hbar/2e$ is the
flux quantum. $E_{J}$ and $E_{c}$ satisfy $\Xi\gg E_{J}\sim E_{c}\gg k_{B}T$, where $\Xi$ is superconducting energy gap, $k_{B}T$
denotes the energy of thermal excitation \cite{Sci296886,Pt5842}.
The Hamiltonian within the basis of Cooper-pair number
states $\{|n\rangle,|n+1\rangle\}$ for the system reads
\begin{align}\label{eq3a-1}
  H_{0}=\sum_{n}{[E_{c}(n-n_{d})^2|n\rangle \langle n|-\frac{E_{J0}}{2}(|n\rangle\langle n+1|+H.c.)]}.
\end{align}
where $E_{c}=2e^{2}/C_{t}$ is the charging energy scale, $C_{t}$ is the total capacitance of the box, and $n_{d}=C_{g}V_{d}/2e$ indicates
the induced gate charges.
According to Refs. \cite{Pc492138,Oc364185,Pra96023843}, we select the lowest level states $|g\rangle$, $|e\rangle$, and $|a\rangle$ to apply the present STA method, which can be
expanded in terms of Cooper-pair states $|n\rangle$ as $|k\rangle=\sum_{n}c_{kn}|n\rangle$ ($k=g,e,a$). The influence of
population occupied by the fourth level state $|f\rangle$ on the coherent transfer between
$|g\rangle$ and $|e\rangle$ of interest can be neglected because of the weak level anharmonicity has been demonstrated in Ref. \cite{Pra96023843}.
Such a quantum circuit has the well-separated level structure and then
can be considered as an effective artificial atom.

We apply an ac microwave driving $\tilde{V}_{c}=V_{c}\cos{(\omega_{c}t)}$ with a frequency $\omega_{c}$ to the considered CPB to induce the
transition between $|g\rangle$ and $|e\rangle$.
However, because of the weak level anharmonicity, the transitions $|e\rangle\leftrightarrow|a\rangle$ can be also triggered by $\tilde{V}_{c}$.
We refer to $|e\rangle\leftrightarrow|a\rangle$ as a quantum leakage in this paper, and we would like to illustrate the dependence of the leakage on level
harmonicity. Assume that two different microwaves fields
$\tilde{V}_{c}=V_{c}\cos{(\omega_{c}t)}$ and $\tilde{V}_{c}'=V_{c}'\cos{(\omega_{c}'t)}$ are applied to the atom, where $\omega_{c}=\omega_{ea}$ and $\omega_{c}'=\omega_{c}-\delta$,
with $\delta$ being an adjust able variable. The microwaves fields $\tilde{V}_{c}=V_{c}\cos{(\omega_{c}t)}$ and $\tilde{V}_{c}'=V_{c}'\cos{(\omega_{c}'t)}$
induce the resonant transitions $|e\rangle\leftrightarrow|a\rangle$ and $|e\rangle\leftrightarrow|g\rangle$, respectively.
According to Ref. \cite{Oc364185}, consider the initial state is $|e\rangle$, with suitable parameters, for a relatively large detuning $\delta/\Omega_{ea}>6$, one gets the
average occupied probability of level state $|g\rangle$: $\overline{P}_{g}\leq 1\%$. This result demonstrates that the sufficient level anharmonicity
between $\omega_{ea}$ and $\omega_{eg}$ can guarantee the negligible leakage $|e\rangle\leftrightarrow|g\rangle$ induced by $\tilde{V}_{c}$.

Then, we address how to realize robust population transfer when the leakage errors are negligible.
The interaction Hamiltonian between the microwave pulse $\tilde{V}_{c}$ and the CPB system reads
\begin{align}\label{eq3a-2}
  H_{cs}=-2E_{c}\tilde{n}_{c}\sum_{n}(n-n_{g})|n\rangle\langle n|,
\end{align}
where $\tilde{n}_{c}=n_{c}\cos{(\omega_{c}t)}$, $n_{c}=C_{g}V_{c}/2e$.
The transition matrix element between $|e\rangle$ and $|a\rangle$ is
\begin{align}\label{eq3a-3}
  t_{ea}=\langle e|H_{cs}|a\rangle=-2E_{c}\tilde{n}_{c}\sum_{n}(n-n_{g})c_{en}^{*}c_{an}=\Omega_{ea}\cos{(\omega_{c}t)}.
\end{align}
Owing to the prohibition by the parity-symmetry determined selection rule, the electric
interaction with a diagonal coupling form does not cause
the transition between $|g\rangle$ and $|a\rangle$ \cite{Pra96023843}.
However, allowed by the level-transition rule, the magnetic interaction Hamiltonian
\begin{align}\label{eq3a-4}
  H_{cp}=-\frac{E_{Jp}}{2}\sum_{n}(|n\rangle\langle n+1|+H.c.),
\end{align}
can give rise to the wanted coupling between $|g\rangle$ and $|a\rangle$. The transition matrix element between $|g\rangle$ and $|a\rangle$ is
\begin{align}\label{eq3a-5}
  t_{ga}=\langle g|H_{cs}|a\rangle=-\pi E_{Jp}\frac{\Phi_{p}}{\Phi_{0}}\sin{(\pi\frac{\Phi_{d}}{\Phi_{0}})}O_{ga}=\Omega_{p}\cos{(\omega_{p}t)},
\end{align}
where $O_{ga}=\sum_{n,m}c_{gn}^{*}c_{am}\langle n|(|n\rangle\langle n+1|+H.c.)|m\rangle$.
Hence, by applying the two microwave drivings $\tilde{\Phi}_{p}$ and $\tilde{V}_{s}$, the interaction of $\Lambda$-configuration, given in Fig. \ref{fig0} (b), can be realized.
The corresponding Hamiltonian is described by
\begin{align}\label{eq3-3}
  H_{0}(t)=&\frac{\hbar}{2}[\Omega_{p}(t)|a\rangle\langle g|+\Omega_{s}(t)|a\rangle\langle e|+H.c.]\cr
            &+\hbar\Delta_{1}(t)|a\rangle\langle a|+\hbar\Delta_{2}(t)|e\rangle\langle e|,
\end{align}
under the rotating wave approximation (RWA).
Here $\Omega_{p}(t)=\Omega_{ga}$ and $\Omega_{s}(t)=\Omega_{ea}$ (chosen real for simplicity) are the
pump and Stokes Rabi frequencies coupling the transitions $|g\rangle\leftrightarrow|a\rangle$ and
$|e\rangle\leftrightarrow|a\rangle$, respectively. $\Delta_{1}(t)=\Delta_{ga}$ and $\Delta_{2}(t)=\Delta_{ea}-\Delta_{ga}$ are the detunings.
Here we consider off-resonant couplings that $\Delta_{ga}=(E_{a}-E_{g})/\hbar-\omega_{p}$ and $\Delta_{ea}=(E_{a}-E_{e})/\hbar-\omega_{c}$.

To apply STA method for a accelerated population transfer, according to the conditions:
(i) the functional form of the evolution path $|\phi_{0}(t)\rangle$
should be more simple than that of its orthogonal partners;
(ii) the elements in the evolution path $|\phi_{0}(t)\rangle$ are better to be nonzero,
the evolution path can be designed by choosing $|\phi_{0}(t)\rangle=|\zeta_{3}^{(3)}\rangle$.
For the sake of convenience and to connect with the previous works \cite{Pra86033405,Pra95062319}, we set parameters $\varphi_{1}=\chi_{2}$, $\varphi_{2}=\chi_{1}+\chi_{2}+\pi$, $\theta=\pi/2-\theta_{1}$, and $\gamma=\pi/2-\theta_{2}$, then we have
\begin{align}\label{eq3-1}
  |\phi_{0}(t)\rangle=\cos{\theta}\cos{\gamma}e^{i\varphi_{1}}|g\rangle+\sin{\gamma}|a\rangle
                      +\sin{\theta}\cos{\gamma}e^{i\varphi_{2}}|e\rangle,
\end{align}
where $\theta$, $\gamma$, and $\varphi_{1,(2)}$ are time-dependent parameters.
Then, its orthogonal partners could be chosen as
\begin{align}\label{eq3-2}
  |\phi_{1}(t)\rangle=&-\frac{1}{\sqrt{2}}[(\sin{\gamma}\cos{\theta}+i\sin{\theta})e^{i\varphi_{1}}|g\rangle-\cos{\gamma}|a\rangle\cr
                      &+(\sin{\gamma}\sin{\theta}-i\cos{\theta})e^{i\varphi_{2}}|e\rangle], \cr
  |\phi_{2}(t)\rangle=&-\frac{1}{\sqrt{2}}[(\sin{\gamma}\cos{\theta}-i\sin{\theta})e^{i\varphi_{1}}|g\rangle-\cos{\gamma}|a\rangle\cr
                      &+(\sin{\gamma}\sin{\theta}+i\cos{\theta})e^{i\varphi_{2}}|e\rangle].
\end{align}
To satisfy the condition given in Eq. (\ref{eq1-6}), by substituting Eqs. (\ref{eq3-3}) and (\ref{eq3-1}) into $\langle 2|H_{1}(t)|1\rangle$ and $\langle 3|H_{1}(t)|1\rangle$, we have
\begin{widetext}
\begin{align}\label{eq3-5a}
  \text{Re}[\langle 2|RH_{0}(t)R^{\dag}|1\rangle]=&\frac{\hbar}{2\sqrt{2}}[\Delta_{1}\sin{2\gamma}-\Delta_{2}\sin^{2}{\theta}\sin{2\gamma} \cr
                                                             &+\Omega_{p}(\cos{\theta}\cos{\varphi_{1}}\cos{2\gamma}+\sin{\theta}\sin{\varphi_{1}}\sin{\gamma})\cr
                                                             &+\Omega_{s}(\sin{\theta}\cos{\varphi_{2}}\cos{2\gamma}-\cos{\theta}\sin{\varphi_{2}}\sin{\gamma})],
\end{align}
\begin{align}\label{eq3-5b}
  \text{Im}[\langle 2|RH_{0}(t)R^{\dag}|1\rangle]=&\frac{\hbar}{2\sqrt{2}}[-\Delta_{2}\cos{\gamma}\sin{2\theta}
                                                                       +\Omega_{p}(\cos{\theta}\sin{\varphi}+\sin{\gamma}\sin{\theta}\cos{\varphi})\cr
                                                                      &+\Omega_{s}(\sin{\theta}\sin{\varphi_{2}}-\sin{\gamma}\cos{\theta}\cos{\varphi_{2}})],
\end{align}
\begin{align}\label{eq3-5c}
  \text{Re}[\langle 3|RH_{0}(t)R^{\dag}|1\rangle]=&\frac{\hbar}{2\sqrt{2}}[\Delta_{1}\sin{2\gamma}-\Delta_{2}\sin^{2}{\theta}\sin{2\gamma} \cr
                                                             &+\Omega_{p}(\cos{\theta}\cos{\varphi_{1}}\cos{2\gamma}-\sin{\theta}\sin{\varphi_{1}}\sin{\gamma})\cr
                                                             &+\Omega_{s}(\sin{\theta}\cos{\varphi_{2}}\cos{2\gamma}+\cos{\theta}\sin{\varphi_{2}}\sin{\gamma})],
\end{align}
\begin{align}\label{eq3-5d}
  \text{Im}[\langle 3|RH_{0}(t)R^{\dag}|1\rangle]=&\frac{\hbar}{2\sqrt{2}}[\Delta_{2}\cos{\gamma}\sin{2\theta}
                                                                      +\Omega_{p}(\cos{\theta}\sin{\varphi}-\sin{\gamma}\sin{\theta}\cos{\varphi}) \cr
                                                                      &+\Omega_{s}(\sin{\theta}\sin{\varphi_{2}}+\sin{\gamma}\cos{\theta}\cos{\varphi_{2}})],
\end{align}
\begin{align}\label{eq3-5e}
  \text{Re}[\langle 2|iR(\partial_{t}R^{\dag})|1\rangle]=&\frac{1}{\sqrt{2}}\cos{\gamma}(\dot{\theta}
                                                                      +\dot{\varphi}_{1}\sin{\gamma}\cos^{2}{\theta}+\dot{\varphi}_{2}\sin{\gamma}\sin^{2}{\theta}),
\end{align}
\begin{align}\label{eq3-5f}
  \text{Im}[\langle 2|iR(\partial_{t}R^{\dag})|1\rangle]=&\frac{1}{2\sqrt{2}}\sin{2\theta}\cos{\gamma}(\dot{\varphi}_{2}-\dot{\varphi}_{1})+\frac{1}{\sqrt{2}}\dot{\gamma},
\end{align}
\begin{align}\label{eq3-5g}
  \text{Re}[\langle 3|iR(\partial_{t}R^{\dag})|1\rangle]=&\frac{1}{\sqrt{2}}\cos{\gamma}(-\dot{\theta}
                                                                      +\dot{\varphi}_{1}\sin{\gamma}\cos^{2}{\theta}+\dot{\varphi}_{2}\sin{\gamma}\sin^{2}{\theta}),
\end{align}
\begin{align}\label{eq3-5h}
  \text{Im}[\langle 3|iR(\partial_{t}R^{\dag})|1\rangle]=&\frac{1}{2\sqrt{2}}\sin{2\theta}\cos{\gamma}(\dot{\varphi}_{1}-\dot{\varphi}_{2})+\frac{1}{\sqrt{2}}\dot{\gamma},
\end{align}
\end{widetext}
where Re$[\cdot]$ and Im$[\cdot]$ denote the real and imaginary parts of argument, respectively.
The equations $\langle 2|H_{1}(t)|1\rangle=0$ and $\langle 3|H_{1}(t)|1\rangle=0$ ask for
\begin{align}\label{eq3-5i}
  \text{Re}[\langle 2|RH_{0}(t)R^{\dag}|1\rangle]&=\text{Re}[\langle 2|iR(\partial_{t}R^{\dag})|1\rangle], \cr
  \text{Im}[\langle 2|RH_{0}(t)R^{\dag}|1\rangle]&=\text{Im}[\langle 2|iR(\partial_{t}R^{\dag})|1\rangle],
\end{align}
and
\begin{align}\label{eq3-5j}
  \text{Re}[\langle 3|RH_{0}(t)R^{\dag}|1\rangle]&=\text{Re}[\langle 3|iR(\partial_{t}R^{\dag})|1\rangle], \cr
  \text{Im}[\langle 3|RH_{0}(t)R^{\dag}|1\rangle]&=\text{Im}[\langle 3|iR(\partial_{t}R^{\dag})|1\rangle],
\end{align}
respectively.
Then, solving Eqs. (\ref{eq3-5i}) and (\ref{eq3-5j}) shows,
\begin{align}\label{eq3-6}
  \Omega_{p}(t)=&\frac{2}{\sin{\varphi_{1}}}(\dot{\theta}\cot{\gamma}\sin{\theta}+\dot{\gamma}\cos{\theta}), \cr
  \Omega_{s}(t)=&\frac{2}{\sin{\varphi_{2}}}(-\dot{\theta}\cot{\gamma}\cos{\theta}+\dot{\gamma}\sin{\theta}), \cr
  \Delta_{1}(t)=&-\frac{\cot{2\gamma}}{2}[\Omega_{p}(t)\cos{\theta}\cos{\varphi_{1}}+\Omega_{s}(t)\sin{\theta}\cos{\varphi_{2}}] \cr
                 &+\dot{\varphi}_{1}\cos^{2}{\theta}+\dot{\varphi}_{2}\sin^{2}{\theta}+\Delta_{2}(t)\sin^{2}{\theta}, \cr
  \Delta_{2}(t)=&[\frac{\Omega_{p}(t)\cos{\varphi_{1}}}{2\cos{\theta}}-\frac{\Omega_{s}(t)\cos{\varphi_{2}}}{2\sin{\theta}}]\tan{\gamma} \cr
                &+\dot{\varphi}_{1}-\dot{\varphi}_{2}.
\end{align}
The solution for the evolution equation $i\hbar \partial_{t}|\psi_{0}(t)\rangle=H_{0}(t)|\psi_{0}(t)\rangle$ is
$|\psi_{0}(t)\rangle=e^{i\beta_{0}}|\phi_{0}(t)\rangle$, with
\begin{align}\label{eq3-7}
  \beta_{0}=-\int_{t_{i}}^{t}[\dot{\varphi}_{1}+(\dot{\theta}\tan{\theta}+\dot{\gamma}\tan{\gamma})\cot{\varphi_{1}}]dt'.
\end{align}
For the sake of simplification, we might choose $\varphi_{1}=-\varphi_{2}=\varphi=\text{const}$ and $0<\varphi<\pi/2$.
Thus,
\begin{align}\label{eq3-7a}
  \Delta_{1}(t)=&-2\cot{\varphi}[(\dot{\theta}\cot{\gamma}\sin{2\theta}+\dot{\gamma}\cos{2\theta})\cot{2\gamma}\cr
               &+(\dot{\theta}\cot{2\theta}-\dot{\gamma}\tan{\gamma})\sin^{2}{\theta}],\cr
  \Delta_{2}(t)=&-2\cot{\varphi}(\dot{\theta}\cot{2\theta}-\dot{\gamma}\tan{\gamma}).
\end{align}
Obviously, by choosing $\varphi=\pi/2$, we have $\Delta_{1}=\Delta_{2}=0$.
The pulses, for convenience, can be expressed as
\begin{align}\label{eq3-7b}
  \Omega_{p}(t)=\Omega_{0}(t)\sin{\tilde{\theta}}, \ \Omega_{s}(t)=\Omega_{0}(t)\cos{\tilde{\theta}},
\end{align}
where
\begin{align}\label{eq3-7c}
  {\Omega}_{0}(t)=&\frac{2}{\sin{\varphi}}\sqrt{\dot{\theta}^{2}\cot^{2}{\gamma}+\dot{\gamma}^{2}},\cr
  \tilde{\theta}=&\theta+\arctan{(\frac{\dot{\gamma}}{\dot{\theta}\cot{\gamma}})}.
\end{align}

If the goal is to drive a system from an initial state $|g\rangle$ to a target state $|e\rangle$,
and in order to simulate the pulses with a finite duration,
the boundaries for the parameters $\theta$ and $\gamma$ should be
\begin{align}\label{eq3-8}
  \theta(t_{i})&=0, \ \theta(t_{f})=\pi/2, \cr
  \gamma(t_{i})&=0, \ \gamma(t_{f})=0, \cr
  \dot{\gamma}(t_{i})&=0,\ \dot{\gamma}(t_{f})=0.
\end{align}
To satisfy these boundaries, we choose Vitanov function for $\theta$ and Gaussian function for $\gamma$ as
\begin{align}\label{eq3-9}
  \theta=\frac{\pi}{2(1+e^{-t/\tau_{1}})}, \ \gamma=\gamma_{0}e^{-t^{2}/\tau_{2}^{2}},
\end{align}
with $0<\tau_{1}<0.12T$, $0.2T<\tau_{2}<0.3T$ ($T=t_{f}-t_{i}$ denotes the total interaction time),
and $0<\gamma_{0}<0.5\pi$ decides the maximal population for state $|a\rangle$.
In experiment, the shapes of the driving pulses with these parameters can be
modulated by electrooptic modulators \cite{Natc712479,arxiv1607}.

\begin{figure}
 \scalebox{0.13}{\includegraphics {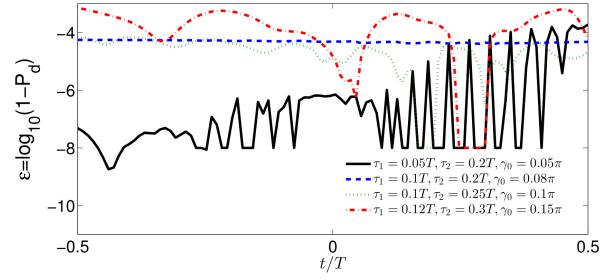}}
 \caption{
         Logarithmic scale of the deviation from a perfect evolution process along the path $|\phi_{0}(t)\rangle$ with random
         parameters within the selectable range for the three-level system.
         }
 \label{fig1}
\end{figure}

\begin{figure}
 \scalebox{0.13}{\includegraphics {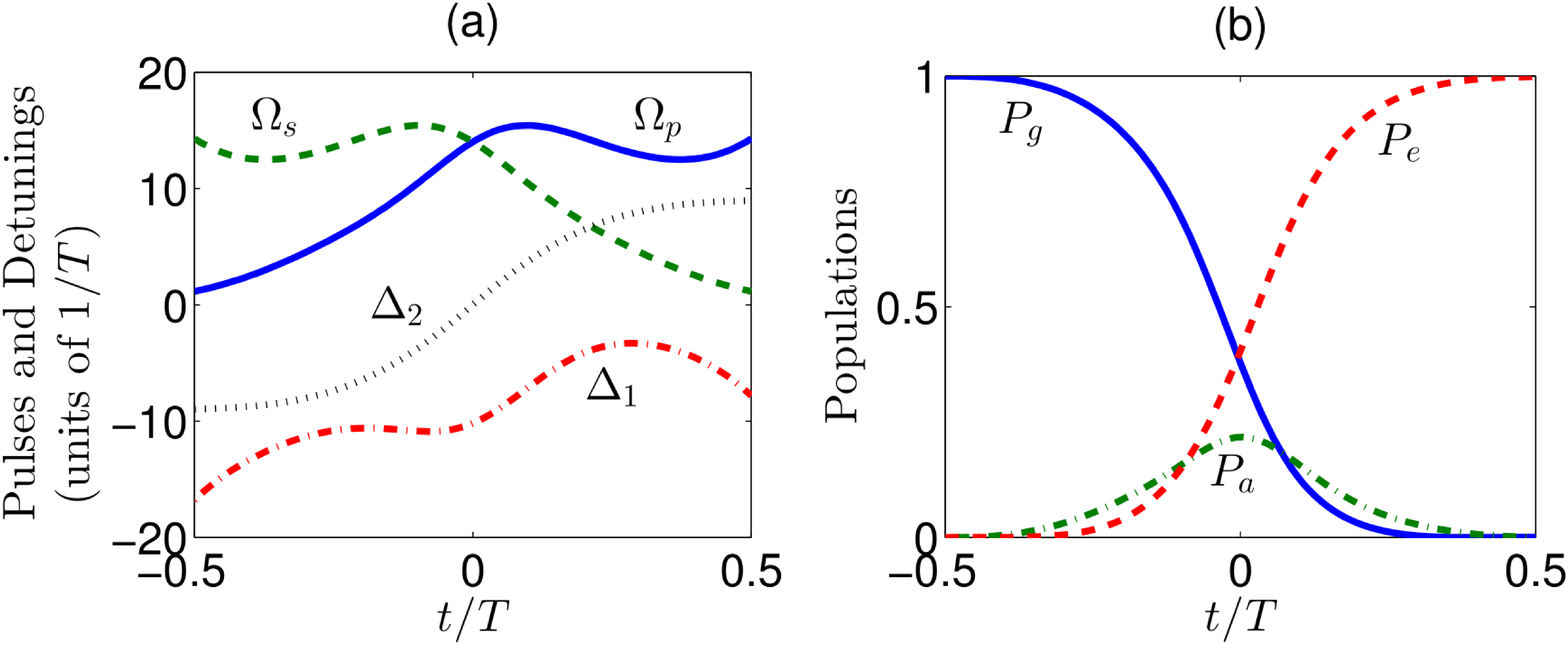}}
 \caption{
         (a) The designed Rabi frequencies and detunings [Eq. (\ref{eq3-6})] versus time.
         (b) The ultrafast population transfer governed by Hamiltonian ${H}_{0}(t)$ in Eq. (\ref{eq3-3}).
         Parameters for Eqs. (\ref{eq3-7a}) and (\ref{eq3-7b}) are $\tau_{1}=0.115T$, $\tau_{2}=0.3T$, $\gamma_{0}=0.15\pi$, and $\varphi=\pi/4$.
         Choosing $\Omega_{0}=0.16\times 2\pi$GHz according to Ref. \cite{Pra96023843}, the time required to reach the target state $|e\rangle$ is only $T\approx16$ns which is much shorter than
         $T=46$ns mentioned in Ref. \cite{Pra96023843}.
         }
 \label{fig3}
\end{figure}

\begin{figure}
 \scalebox{0.13}{\includegraphics {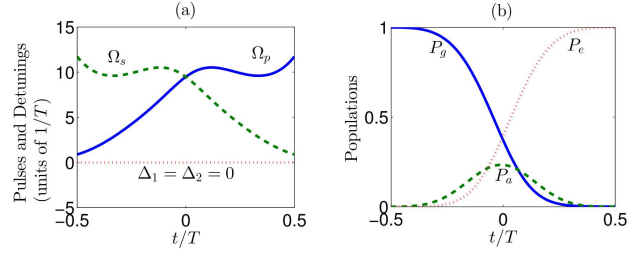}}
 \caption{
         (a) The designed Rabi frequencies and detunings [Eq. (\ref{eq3-6})] versus time when $\varphi=\pi/2$.
         (b) The ultrafast population transfer governed by Hamiltonian ${H}_{0}(t)$ in Eq. (\ref{eq3-3}).
         Parameters are $\tau_{1}=0.115T$, $\tau_{2}=0.3T$, and $\gamma_{0}=0.15\pi$.
         }
 \label{fig3a}
\end{figure}

First of all, we would like to verify whether the system evolves along the path as expected or not.
We define an error function $\varepsilon=\log_{10}[1-P_{d}(t)]$ for analysis, where $P_{d}(t)=|\langle\phi_{0}(t)|\psi_{0}(t)\rangle|^{2}$.
As shown in Fig. \ref{fig1}, within the selectable range for the parameters,
we verify with an accuracy to about three digits that the system has been
driven exactly along the path as expected.
Then with parameters $\{\gamma_{0}=0.15\pi,\ \tau_{1}=0.115T,\ \tau_{2}=0.3T, \ \varphi=\pi/4\}$,
we display the parameters [$\Omega_{p,(s)}(t)$ and $\Delta_{1,(2)}(t)$] and time-dependent populations
(marked as $P_{n}$ for state $|n\rangle$) as an example in Figs. \ref{fig1} (a) and (b), respectively.
Shown in the figure, a nearly perfect population transfer from the initial state $|g\rangle$
to the target state $|e\rangle$ could be obtained with the final population for state $|e\rangle$
is $P_{e}(t_{f})=0.9997$. Generally speaking, time-dependent detunings are relatively harder to experimentally realize than time-independent ones.
For the present scheme, according to Eq. (\ref{eq3-7a}), when we choose $\varphi=\pi/2$, we have $\Delta_{1}=\Delta_{2}=0$. The corresponding
time-dependent parameters [$\Omega_{p,(s)}(t)$ and $\Delta_{1,(2)}(t)$] and populations are shown in Fig. \ref{fig3a}. Also, a nearly perfect population transfer
with final population $P_{e}(t_{f})=0.9995$ can be achieved. Contrasting Fig. \ref{fig3a} with Fig. \ref{fig3}, the total evolution time
required in the resonant case ($\varphi=\pi/2$) is shorter than that in the off-resonant case ($\varphi=\pi/4$).


\begin{figure}
 \scalebox{0.13}{\includegraphics {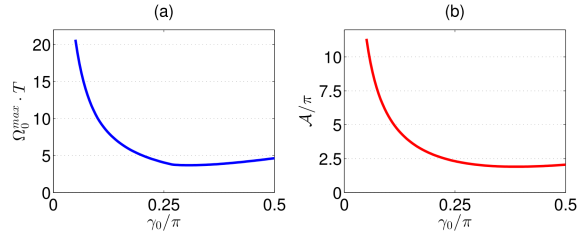}}
 \caption{
         (a) The relationship between total interaction time scale $T\Omega_{0}^{max}$ and $\gamma_{0}$.
         (b) The relationship between total pulse area $\mathcal{A}$ and $\gamma_{0}$.
         Parameters are chosen as $\tau_{1}=0.12T$, $\tau_{2}=0.3T$, and $\varphi=\pi/2$.
         }
 \label{fig2}
\end{figure}

For convenience, we define a dimensionless parameter $T\Omega_{0}^{max}$
as a measurement scale for total interaction time in the following discussion,
where $\Omega_{0}^{max}$ denotes the maximum value of $\Omega_{0}$.
Beware that ${\Omega}_{0}^{max}$ is usually a little larger than the maximum value for ${\Omega}_{p,(s)}(t)$,
the total interaction time measured by the $T{\Omega}_{0}^{max}$ is in fact a little larger than the real one.
While, $T{\Omega}_{0}^{max}$ would help a lot for quantitative analysis in the total interaction time, so, we tend to use $T{\Omega}_{0}^{max}$
as a measurement scale for the total interaction time.
Substituting Eq. (\ref{eq3-9}) into Eq. (\ref{eq3-7c}), we can find
the pulse maximum amplitude $\Omega_{0}^{max}$ is obviously in inverse proportion to $\tau_{1,(2)}$.
That is, $\tau_{1,(2)}$ should be chosen as large as possible,
i.e., $\tau_{1}=0.12T$ and $\tau_{2}=0.3T$, to shorten the interaction time for the process.
The pulse area defined as
$\mathcal{A}=\int_{-\infty}^{+\infty}dt\sqrt{\Omega_{p}^{2}+\Omega_{s}^{2}}$ which
is used to measure the total energy cost of the quantum process,
in this case, is given as
\begin{align}\label{eq3-10}
  \mathcal{A}=\frac{2}{\sin{\varphi}}\int_{-\infty}^{+\infty}dt\sqrt{\dot{\theta}^{2}\cot^{2}{\gamma}+\dot{\gamma}^{2}}.
\end{align}
Then, with $\{\tau_{1}=0.12T,\ \tau_{2}=0.3T,\ \varphi=\pi/2\}$, we display $T\Omega_{0}^{max}$ versus $\gamma_{0}$ and $\mathcal{A}$
versus $\gamma_{0}$ in Figs. \ref{fig2} (a) and (b), respectively. Shown in the figure,
both $T\Omega_{0}^{max}$ and $\mathcal{A}$, in general, decrease with the increasing of $\gamma_{0}$,
while when $\gamma_{0}>0.3\pi$, $T\Omega_{0}^{max}$ and $\mathcal{A}$ stop decreasing but
increasing very slowly with the increasing of $\gamma_{0}$. When $\gamma_{0}=0.3\pi$, we have
$T\Omega_{0}^{max}\approx 3.696$ and $\mathcal{A}\approx 1.907\pi$ which are,
respectively, the shortest interaction time and the smallest pulse area based on the present method.
We know the naive simplest way to completely transfer the population between
the two ground states in a $\Lambda$-type system without coupling the ground states is two successive $\pi$ pulses, one for
each transition, leading to a total pulse area of $\mathcal{A}=2\pi$ \cite{Aps58243},
and the minimum area for such a process, is $\sqrt{3}\pi$, which corresponds to the singular-Riemannian geodesic \cite{Jmp432107}.
That is, the total interaction time $T\Omega_{0}^{max}$ and the pulse area $\mathcal{A}$ in present method,
are small enough for us to say the population transfer is ultrafast.

Here for comparison, we would like to discuss a situation when $\gamma\rightarrow\text{const}=\gamma_{0}$.
Under such hypothesis, to ensure the system is initially in the path $|\phi_{0}(t)\rangle$,
the error function $\varepsilon=\log_{10}[1-P_{d}(t_{i})]\leq-3$, leading to
$\cos^{2}{\gamma_{0}}\geq0.999\Rightarrow\gamma_{0}\leq10^{-2}\pi$,
should be satisfied. Then, we find $\mathcal{A}={\pi\cot{\gamma_{0}}}/\sin{\varphi}\geq10\pi$.
This is an interesting result because it figures out the minimum pulse area
required for an ideal stimulated Raman adiabatic passage with dark-state evolution.
When $\gamma\rightarrow\text{const}$ and $\varphi=\pi/2$, the vectors $|\phi_{n}(t)\rangle$ ($n=0,1,2$)
are found to be the eigenstates of $H_{0}(t)$ with eigenenergies
$E_{0}=0$, $E_{1}(t)=-E_{2}(t)=\dot{\theta}\cot{\gamma_{0}}$, respectively.
The adiabatic condition $|\langle\phi_{0}(t)|\partial_{t}\phi_{1,(2)}(t)\rangle|\ll |E_{1,(2)}(t)|\Rightarrow\sqrt{2}\cot{\gamma}\gg1$
has been checked to be ideally satisfied. The pulse maximum amplitude,
with $\theta$ in form of Eq. (\ref{eq3-9}), is $\Omega_{0}^{max}=\pi\cot{\gamma_{0}}/(4\tau_{1})$.
For an adiabatic process, by choosing $\gamma_{0}=10^{-2}\pi$ and $\tau_{1}=0.12T$,
we find $T\Omega_{0}^{max}\approx65\pi$ is much larger than that of the present STA method.

\section{Robustness against noise}
\begin{figure}
 \scalebox{0.2}{\includegraphics {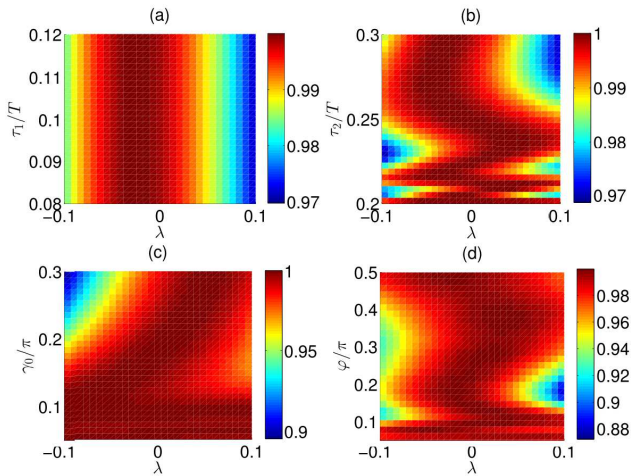}}
 \caption{
         The final population $P_{e}(t_{f})$ for the target state $|e\rangle$
         versus systematic noise. Shown in (a) and (b), relatively speaking, the changes of $\tau_{1}$ and $\tau_{2}$
         affect slightly to the systematic-error sensitivity. Shown in (c), with parameters
         $\{\tau_{1}=0.12T,\ \tau_{2}=0.3T,\ \varphi=\pi/2 \}$, $\gamma_{0}$ should be chosen relatively small to restrain the systematic noise.
         Shown in (d), the systematic-error sensitivity
         decreases with the increasing of $\varphi$. Parameters for (a)-(d)
         are $\{\tau_{2}=0.3T,\ \gamma_{0}=0.15\pi,\ \varphi=\pi/2 \}$, $\{\tau_{1}=0.12T,\ \gamma_{0}=0.15\pi,\ \varphi=\pi/2 \}$,
         $\{\tau_{1}=0.12T,\ \tau_{2}=0.3T,\ \varphi=\pi/2 \}$, and $\{\tau_{1}=0.12T,\ \tau_{2}=0.3T,\ \gamma_{0}=0.15\pi \}$, respectively.
         }
 \label{fig4}
\end{figure}

To check the robustness of the system, we first consider the influence on the fidelity of systematic errors.
Let the ideal, unperturbed Hamiltonian be $H_{0}(t)$. When systematic errors are considered,
the actual, experimentally implemented Hamiltonian is $H_{0s}(t)=H_{0}(t)+\lambda H_{s}(t)$, but the evolution of the pure quantum state is still described
by the Schr\"{o}dinger equation,
\begin{align}\label{eq3-11}
  i\hbar\partial_{t}|\psi(t)\rangle=[H_{0}(t)+\lambda H_{s}(t)]|\psi(t)\rangle.
\end{align}
We assume the errors affect the Rabi frequencies $\Omega_{p}(t)$ and $\Omega_{s}(t)$ but not the deuntings $\Delta_{1,(2)}(t)$.
The error Hamiltonian can be assumed as in form of
\begin{align}\label{eq3-12}
  H_{s}(t)=\frac{\hbar}{2}[\Omega_{p}(t)|a\rangle\langle g|+\Omega_{s}(t)|a\rangle\langle e|]+H.c..
\end{align}
By numerical simulation, we show the final population $P_{e}(t_{f})$ for the target state $|e\rangle$
versus systematic noise $\lambda$ in Fig. \ref{fig4}. Relatively speaking, the systematic-error sensitivity changes slightly
with the change of $\tau_{1}$ [see Fig. \ref{fig4} (a)] or $\tau_{2}$ [see Fig. \ref{fig4} (b)],
and the transfer process is relatively less sensitive to systematic error with lager $\tau_{1}$ and $\tau_{2}$ than smaller ones.
The changes of $\gamma_{0}$ and $\varphi$ affect the systematic-error sensitivity of the transfer process
more seriously than those of $\tau_{1}$ and $\tau_{2}$ as shown in Figs. \ref{fig4} (c) and (d), respectively.
With parameters $\{\tau_{1}=0.12T,\ \tau_{2}=0.3T,\ \varphi=\pi/2 \}$, $\gamma_{0}$
should be chosen relatively small to restrain the systematic noise.
The best choice for $\varphi$ to restrain the systematic noise as shown in Fig. \ref{fig4} (d), is $\varphi=\pi/2$.

\begin{figure}
 \scalebox{0.2}{\includegraphics {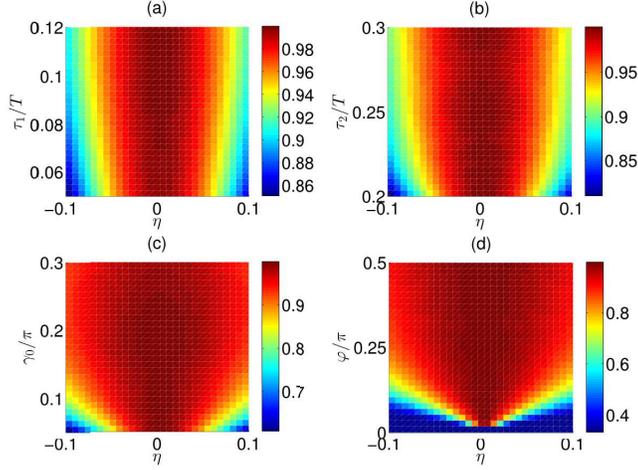}}
 \caption{
         The final population $P_{e}(t_{f})$ for the target state $|e\rangle$
         versus amplitude noise. The sensitivity with respect to amplitude-noise error
         obviously decreases with the increasing of each of the four parameters $\{\tau_{1},\ \tau_{2},\ \gamma_{0}, \varphi\}$.
         Parameters for (a)-(d)
         are $\{\tau_{2}=0.3T,\ \gamma_{0}=0.15\pi,\ \varphi=\pi/2 \}$, $\{\tau_{1}=0.12T,\ \gamma_{0}=0.15\pi,\ \varphi=\pi/2 \}$,
         $\{\tau_{1}=0.12T,\ \tau_{2}=0.3T,\ \varphi=\pi/2 \}$, and $\{\tau_{1}=0.12T,\ \tau_{2}=0.3T,\ \gamma_{0}=0.15\pi \}$, respectively.
         }
 \label{fig5}
\end{figure}

The second type of error is a stochastic one, which means the Hamiltonian is perturbed by
some stochastic part $\eta H_{a}(t)$ describing the amplitude noise. The Schr\"{o}dinger equation in
the Stratonovich sense reads
\begin{align}\label{eq3-13}
  i\hbar\partial_{t}|\psi(t)\rangle=[H_{0}(t)+\eta H_{a}(t)\xi(t)]|\psi(t)\rangle,
\end{align}
where $\eta$ is the strength of the amplitude noise and $\xi(t)=\frac{d W_{t}}{dt}$
is heuristically the time derivative of the Brownian motion $W_{t}$. $\xi(t)$ should
satisfy $\langle\xi(t)\rangle=0$ and $\langle\xi(t)\xi(t')\rangle=\delta(t-t')$
because the noise should have zero mean and should be uncorrelated at different times.
Beware that the evolution of the quantum state with amplitude noise can only be described by a master
equation \cite{Njp14093040,CHJ99}. The dynamical evolution described by Eq. (\ref{eq3-13}) is in fact inaccurate.
According to Ref. \cite{Njp14093040}, when different realizations are averaged over,
the density operator $\rho(t)$ should satisfy
\begin{align}\label{eq3-14}
  \partial_{t}\rho(t)=-\frac{i}{\hbar}[H_{0}(t),\rho(t)]
                      -\frac{\eta^2}{2\hbar^2}[H_{a}(t),[H_{a}(t),\rho(t)]].
\end{align}
In this paper, we consider independent amplitude noise in $\Omega_{p}(t)$ as well as in $\Omega_{s}(t)$ with the same intensity $\eta^2$,
then, the master equation is
\begin{align}\label{eq3-15}
  \partial_{t}\rho(t)=&-\frac{i}{\hbar}[H_{0}(t),\rho(t)]-\frac{\eta^2}{2\hbar^2}[H_{a}^{p}(t),[H_{a}^{p}(t),\rho(t)]]\cr
                                                        &-\frac{\eta^2}{2\hbar^2}[H_{a}^{s}(t),[H_{a}^{s}(t),\rho(t)]],
\end{align}
where
\begin{align}\label{eq3-16}
  H_{a}^{p}&=\frac{\hbar}{2}\Omega_{p}(t)|a\rangle\langle g|+H.c., \cr
  H_{a}^{s}&=\frac{\hbar}{2}\Omega_{s}(t)|a\rangle\langle e|+H.c..
\end{align}
Defining the final population for the target state as $P_{e}(t_{f})=|\langle e|\rho(t_{f})|e\rangle|$,
the sensitivity with respect to amplitude-noise error is shown in Fig. \ref{fig5}.
The sensitivity with respect to amplitude-noise error obviously decreases with the increasing of both $\tau_{1}$ and $\tau_{2}$ as
shown in Figs. \ref{fig5} (a) and (b).
The changes of $\gamma_{0}$ and $\varphi$, especially, when they are relatively small,
affect the fidelity of the transfer very seriously in the presence of amplitude-noise errors
as shown in Fig. \ref{fig5} (c) and (d).
The results from Figs. \ref{fig4} and \ref{fig5} drive us to choose
relatively large $\tau_{1,(2)}$, $\gamma_{0}$, and $\varphi$, such as $\{\tau_{1}=0.12T,\ \tau_{2}=0.3T,\ \gamma_{0}=0.15\pi,\ \varphi=\pi/2\}$, so that
the transfer process would be robust against systematic error and amplitude-noise error
.

\section{conclusion}
In conclusion, we have proposed a promising method to
implement STA without additional couplings.
The strategy is to design a nonadiabatic evolution path
which is decoupled from its orthogonal partners.
We focus on designing the evolution path without
making explicit use of a reference Hamiltonian $H_{0}(t)$. In this way,
applying accelerated dynamics to a wider field would be much easier
because there is a tractable algorithm to find orthogonal complete
vectors for arbitrary dimension space (see Sec. II), while there is not a tractable algorithm to find analytical eigenstates for a general Hamiltonian.
As an exemplified case, we apply the present STA method to accelerate population
transfer within a transmon-type qutrit. The qutrit, constituted by the lowest
three levels, can be coupled to the microwave drivings of ac
voltage and time-dependent bias flux. With
the available parameters, population transfer can
be drastically accelerated via the present STA method as demonstrated by numerical simulation.
Numerical simulation also shows that the total pulse area (total energy cost) for the present three-level system is low as $1.9\pi$.
Besides, the accelerated system is robust against systematic and amplitude-noise errors.
We hope that the current work may
open venues for the experimental realization of STA methods in the near future.

The drawback of the present method is that the general expression of nonadiabatic evolution path
is still unclear. In order to ensure the decoupling condition in Eq. (\ref{eq1-6}) is analytically solvable,
the expression of nonadiabatic evolution path should be relatively complex. We hope the future work can overcome this problem.

\section{acknowledgement}

This work was supported by the National Natural Science Foundation
of China under Grants No. 11575045, No. 11374054 and No. 11675046.

\end{document}